\documentclass[graybox,natbib,nosecnum]{svmult}
\bibpunct{(}{)}{;}{a}{}{,} 

\pdfoutput=1   

\usepackage{mathptmx}       
\usepackage{helvet}         
\usepackage{courier}        
\usepackage{type1cm}        
\usepackage{mathtools}		

\usepackage{makeidx}         
\usepackage{graphicx}        
\usepackage{multicol}        
\usepackage[bottom]{footmisc}
\usepackage[normalem]{ulem}	
\usepackage{hyperref}  
\usepackage[dvipsnames]{xcolor}

\usepackage{amssymb}
\usepackage{amsmath}
\usepackage{siunitx}

\usepackage{soul}   

\newcommand{\eps}{\varepsilon}			
\def\t#1{\mathrm{#1}}					
\def\defEq{\coloneqq}

\newcommand{\vk}{v_\mathrm{K}}			

\newcommand{\stokes}{\mathrm{St}}					

\makeindex             

\begin{document}

\title*{Instabilities and Flow Structures in Protoplanetary Disks:
Setting the Stage for Planetesimal Formation}
\titlerunning{Instabilities and Flow Structures in Protoplanetary Disks} 
\author{Hubert Klahr, Thomas Pfeil, Andreas Schreiber}
\institute{H. Hubertus Klahr \at Max Planck Institut for Astronomy, K\"onigstuhl 17, D-69117 Heidelberg, Germany, \email{klahr@mpia.de}}
%
%
\maketitle

\abstract{This chapter highlights the properties of turbulence and meso-scale flow structures in protoplanetary disks and their role in the planet formation process. 
{Here we focus on the formation of planetesimals from a gravitational collapse of a pebble cloud. 
Large scale and long lived flow structures - vortices and zonal flows - are a consequence of weak magneto and hydrodynamic instabilities in the pressure and entropy stratified quasi-Keplerian shear flow interacting with the fast rotation of the disk. The vortices and zonal flows on the other hand are particle traps tapping into the radial pebble flux of the disk, leading to locally sufficient accumulations to trigger gravitational collapse, directly converting pebbles to many kilometer sized planetesimals. This collapse is moderated by the streaming instability, which is a back-reaction from the particle accumulations onto the gas flow. 
Without trapping pebbles and increasing thus the local solid to gas ratio, this back-reaction would ultimately prevent the formation of planetsimals via turbulent diffusion. The formation of long lived flow structures is therefore a necessary condition for an efficient and fast formation of planetesimals.}}
\section{Introduction}
A range of hydrodynamical and magneto-hydro-dynamical instabilities have been demonstrated to operate in disks around young stars. These instabilities contribute to the angular momentum transport in disks and thus enable the accretion of mass onto the central star. This accretion process sets the evolution of the temperature and density structure but also the dynamical structure of a disk during the early stages of planet formation, e.g. before photo evaporation has drained the gas from the disk after 1-10 Myrs. Whereas Earth and the other terrestrial planets will ultimately form after the removal of the gas disk, many crucial steps in planet formation take place in a gaseous environment. These processes range from the early coagulation of ice and dust to pebbles, the formation of planetesimals and cores of planets, to gas accretion onto giants and migration via tidal interaction with the gas disk (see chapters by Armitage, by Andrews and Birnstiel and by Nelson).
All these processes depend delicately on the proper dynamical and thermodynamical state of the surrounding disk. In other words: \textit{in a strictly laminar nebula around a young star, i.e. with a monotonously dropping radial pressure and no other gas motion but the slightly sub-Keplerian rotation, there will hardly be any planets formed and certainly not the ones that we observe in our solar system.} This chapter shall give an overview to these different kinds of instabilities. We discuss their role on shaping the disk structure and providing birthplaces for planetesimals as an early key process for planet formation.

\section{What Makes a Circumstellar Disk?}
\begin{figure}
\includegraphics[scale=.55,trim={0.8cm 0 0 0},clip]{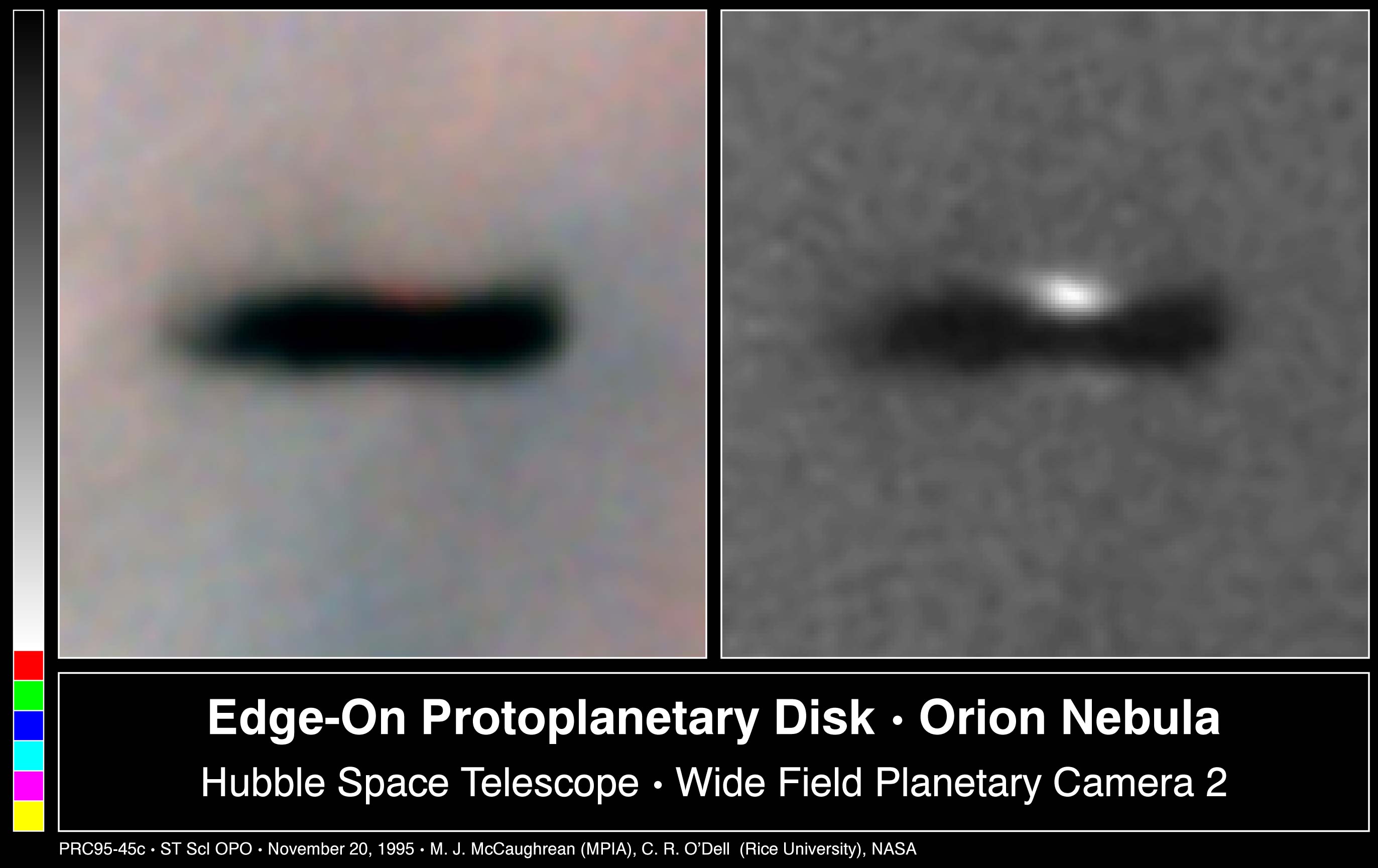}
\includegraphics[scale=.6,trim={0.8cm 0 0 0},clip]{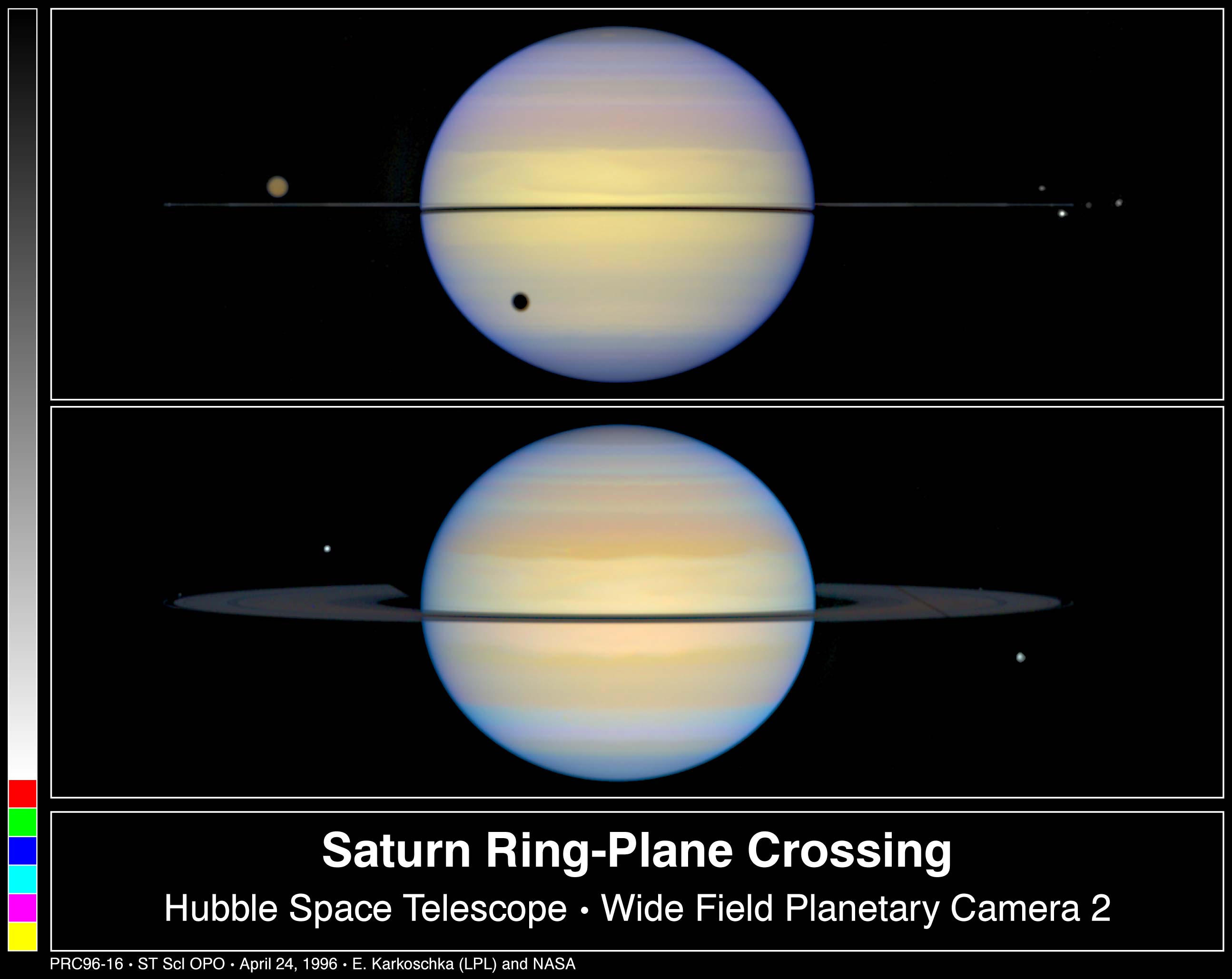}
\caption{Disk in the Orion Nebula vs. the gas free "disk" around Saturn. Compare the thick gaseous disks with the razor thin gas free disk. Image credits for the protoplanetary disk: Mark McCaughrean (Max-Planck-Institute for Astronomy), C. Robert O'Dell (Rice University), and NASA/ESA. For Saturns rings: Erich Karkoschka (University of Arizona Lunar \& Planetary Lab) and NASA/ESA.
\label{fig:OrionSaturn}
}
\end{figure}

A gas disk around a young star is fundamentally different from a gas free disk of ice and dust particles, as for instance the rings of Saturn.
Already by visual inspection of images from Saturn's rings with a disk from the Orion Nebula (see Fig.~\ref{fig:OrionSaturn}), we notice that 
the pure particle disk is extremely thin in comparison to its radial extent (aspect ratios are between $1:300.000$ and $1:30.000.000$), whereas disks around young star are relatively thick with an aspect ratio between $1:20 - 1:10$. In Saturn's rings it is mainly the collisions between the ring particles that set its thickness as well as drive its dynamical evolution (for a recent review on Saturn's rings we suggest \cite{2015chvs.book.....M}). 

In case of a circumstellar disk, the particles are merely the tracer of the gas distribution and its dynamical state. Without gas, the dust would quickly sediment to a thin layer, because any initial inclination of the particles will be quickly damped via collisions, just what happened in Saturn's rings.

Thus, the greater thickness of the dust disk is an indication for the
gas turbulence preventing the dust from sedimentation to a thin mid-plane. This is important to note, because the gas itself, consisting mainly out of molecular Hydrogen and Helium, is largely invisible to the observer. Hence, it is common to derive the disk gas mass from the distribution of grains. In a purely laminar disk, the dust would quickly settle to the mid-plane and
rain out into the star on time scales much shorter than the inferred lifetime of disks around young stars. The observations can
only be explained if there is a stirring process is acting onto the dust particles, i.e. their vertical sedimentation has to be counteracted by turbulent diffusion (see chapter by Andrews and Birnstiel).
 
Now, from the images of disks we can already infer that there is turbulent gas, and that this gas has to be turbulent. Unfortunately there is no direct way to quantify
the absolute amount of gas present nor the turbulent strength, because many factors also influence the
shape of the dust distribution, like the dynamical properties of the ice and dust. Therefore, only complex modelling of the disk together with a model for the dust evolution can derive plausible disk properties (see chapter by Andrews and Birnstiel).

{Recent observations of the dust and pebble distribution in disks around young stars hint at the existence of particle traps in these systems \cite{2013Sci...340.1199V,2016ApJ...821L..16C}. The observation of these traps is a hint that structures like vortices and zonal flows do actually exist in potentially planet forming disks. So our task is now to explain their formation in a pre-planet phase of the disk, by a flow instability. If these vortices and zonal flows could only be triggered by an already existing planet, then it would be hard to argue for their role in forming planetesimals. But in the paradigm of forming planetesimals from a gravitational collapse of pebble clouds one needs the pebble trapping and concentrating effect of vortices and zonal flows before planetesimals and thus planets do form. Therefore we are motivated to study the formation and evolution of traps as a consequence of the radially and vertically stratified disk subject to shear, thermal relaxation and magnetic fields at generally low ionisation rates, e.g. a typical circumstellar disk.}

From the disk dynamical side, we can not prove from first principles what the dominating (magneto-) hydrodynamical instabilities in protoplanetary disk will be, nor
what dynamical and thermodynamic state will develop from them. 
Note that even for well observed phenomena like the atmospheric band structures of Jupiter, there is no consensus on the main driver
for the underlying velocity perturbations, see \cite{2005Natur.438..193H}.
But, we can thoroughly investigate all possible (magneto-) hydro-dynamical
mechanisms and check on their actual occurrence in protoplanetary disks, by embedding them in simulations that cover the observed disk properties (\cite{2014prpl.conf..339T})
as well as in population synthesis simulations for exoplanets (e.g. \cite{2014A&A...567A.121D}). The quality of these simulations 
as measured against the actual findings of disk observations as well as exoplanets is then a measure for the 
quality of the underlying physics, e.g. the instabilities and processes that drive and shape turbulence in disks
around young stars. 
By relating simulated disk dynamics to its consequences for disk and planet observation, we try to falsify our models, as directly measuring the
turbulence so far has rendered to be unsuccessful \citep{2018ApJ...856..117F}. The authors find upper limits for observable turbulent velocities in the TW Hya Disk leading to an estimated viscosity with $\alpha < 0.007$. As we will discuss below, even lower values for $\alpha$ would drive sufficient disk evolution and more importantly would be compatible with flow structure formation.

The original motivation to study hydro-dynamical instabilities and turbulence in disks around compact objects comes from the detection
of substantial mass accretion from the disk onto the central object. These compact objects can be black holes, neutron stars or as
in our case a young star. The general problem here lies in the conservation of angular momentum. Mass first has to be 
stripped of angular momentum before being able to fall onto the central object. This angular momentum transport can be related to
large scale magnetic fields, which anchor in the surrounding material, which then would break the disk by transferring the 
angular momentum via magnetic torques directly out of the system \citep{Turner_2014}. 
These torques are magnetized wind torques, not just Alfven waves as an example. There is also a huge body of theory on this in earlier disk wind literature in the context of protostellar jets / disks \citep{2007prpl.conf..277P}.
This picture finds support by the observation of outflows and jets from many accreting objects, which is a magnetic phenomenon, where twisted field lines accelerate
and focus gas parcels from the surface of the disk. This process is especially important in the early phases of star and disk formation, during the collapse of the molecular cloud core \citep{2014prpl.conf..173L}. 
Winds are highly efficient in stripping angular momentum from a rotating system: It can be shown that only one gas particle for each ten in the disk is needed to drive the accretion flow \citep{1986ApJ...301..571P, 1992ApJ...394..117P}.

A second process to allow for accretion onto the central object is turbulent transport in the disk \citep{Turner_2014}. Here, the angular momentum stays in the disk but is transported outward by viscous shear. In this way, an inner annulus of the disk looses its angular momentum
via friction to its outer neighboring annulus, until the outer disk edge is reached, which is then lifted into a wider orbit. And because
the Keplerian rotation law $\Omega \propto R^{-\frac{3}{2}}$ is sufficiently shallow with radius $R$, and 
thus specific angular momentum $l$ increases radially via,
\begin{equation}
l = \Omega R^2  \propto R^{\frac{1}{2}},
\end{equation}
the disk has radially to spread only a little. Thus, most of the disks angular momentum can reside in some 
low amounts of gas mass at a large separation from the star, whereas most of the gas mass, now stripped from angular momentum, falls onto
the star. 

The viscosity in the case of Saturn's rings that drives the viscous evolution is given by particle collisions. The equivalent in 
a gaseous disk is the molecular viscosity, which is found to be much too low for explaining the observed accretion rates. \cite{Shakura_1973} introduced therefor an effective viscosity that is proportional to the thermal pressure in the disk. This unspecified viscous process is then accounted for the driving of the observed accretion. In their work, they defined an $\alpha$-value 
relating turbulent gas velocities $v_\mathrm{t}$ and Alfv\'en velocity $v_\mathrm{A}$ (which is roughly speaking the velocity at which waves propagate via the effective tension of magnetic field lines) the two possible agents to carry angular momentum to the speed of sound $c_\mathrm{s}$ 
\begin{equation}
\alpha^2 = \frac{v_\mathrm{t}^2 + v_\mathrm{A}^2}{c_\mathrm{s}^2},
\end{equation}
and then define a mixing length, e.g. the thickness of the disk $H$ to 
derive a viscosity:
\begin{equation}
\nu = \alpha c_s H.
\end{equation}

The $\alpha$-value has become famous for expressing unknown effective viscosities for the
quantification of accretion mass fluxes, but also famous as a parameterization for the turbulent strength in protoplanetary disks. The later includes it as a description for turbulent diffusion of dust grains, well coupled to the gas. 
Its value for disks around black holes and neutron stars has been found to be close to unity, indicating for these systems that turbulence is almost super-sonic and that magnetic pressures is close to thermal pressure. For disks around young stars $\alpha$-values need to be lower in order to explain observations.
$\alpha = 10^{-3} - 10^{-2}$ are certainly a good assumption (see also the chapter by Armitage), yet neither theory nor observation can 
nail it further down (See also our comment above on the upper limits for turbulence from observations of the TW-Hydra system \cite{2018ApJ...856..117F}. Especially all the known hydro-dynamical instabilities lead to different angular momentum transport efficiencies,
making a universal $\alpha$-value for all locations and times in disks, as well as independent from disk mass, stellar luminosity, etc. a rather rough tool. Lets now finally consider what can be said on the stability of gas disks around young stars. 

\section{Hydro-Dynamics and Stability Analysis}
The equations of hydro-dynamics can be interpreted as conservation equations for
mass, momentum and energy in a fluid.
For instance mass $m = \rho V$ in a volume $V$ can only change,
if some mass flux $\rho v$ over the volume its surface $A$ occurs. 
Here, $\rho$ is the mean density within the volume and $v$ the mean gas velocity.
The differential equation that is describing how this local density changes, in the case of a compressible medium, is 
\begin{equation}
\partial_t \rho + \nabla \rho v = 0,
\end{equation}
known as continuity equation.

The momentum of the gas $\rho v$ is a conserved quantity, yet internal and external
forces can lead to additional accelerations. Internal effects might be the pressure
gradient in the gas $\nabla p$ and viscous stresses, whereas external effects is 
for instance gravity $g$, magnetic fields, drag forces from embedded objects, etc.
For instance, if gravity enters the (1-D) momentum equation:
\begin{equation}
\partial_t \rho v+ \nabla \rho v . v = - \nabla p + g \rho.
\end{equation}
One can now subtract the continuum equation from this momentum equation, and one derives the more common form:
\begin{equation}
\partial_t v+ v \nabla v = - \frac{1}{\rho}\nabla p + g.
\end{equation}
Please note that velocity $v$ is not a conserved quantity, and  
only by simultaneous solving the continuity equation the momentum can be conserved.

Finally, energy $E$ has to be conserved, too. $E$ is typically representing the total energy, i.e. the sum of thermal energy $e_\mathrm{thermal} $, kinetic energy $\frac{\rho v^2}{2}$ and all other additional 
effects, like gravitational potential, magnetic fields, etc., leading to
\begin{equation}
\partial_t E+ \nabla v \left(E + p\right) = \rho  g.
\end{equation}

One obvious problem in solving these equations is that we have four independent unknowns,
density, thermal energy, pressure and velocity, but only two equations. The loop is
closed by introducing an equation of state to the system, determining pressure
as a function of temperature and density, which can be done using an ideal gas
law. 
\begin{equation}
p = \frac{R_{gas}}{\mu} \rho T.
\end{equation}
and
\begin{equation}
e_\mathrm{thermal} = \frac{p}{\gamma -1}
\end{equation}
Besides the values for the gas constant, the molecular mass $\mu$ and 
the ratio of specific heats, also known as adiabatic index $\gamma$ depend
on the gas mixture and the temperature range considered. For details check (\cite{2014pafd.book.....C}).

Finding solutions to this set of equations and especially determining the properties of the resulting
flow is the holy grail of fluid dynamics. The problem lies in the non-linearity of the equations, which
restrict solutions to simplified configurations, or makes the use of super computers necessary.
Yet, both solution strategies can still not explain from first principles for what parameters a system 
will become non-linear unstable and develop turbulence. That problem has haunted physics
since the late days of the 19th century, when Reynolds and his colleagues laid the fundamentals
for turbulence research.

An important tool in fluid dynamics are stability considerations for specific flows.
The strategy in that endeavor is three-fold: First one has to find an equilibrium state
within the hydro-dynamical equations for the considered system - in our case a gas disk around a young star. 
Then, one applies an infinitesimal perturbation to the system by introducing small variations to the system quantities of state, i.e. pressure, velocity field, etc. These perturbations then can be analyzed for growing modes in the linearized 
system. Finally, one puts the findings to a test in specifically tailored, non-linear simulations, which then study the onset of turbulence and saturation effects for the instability.

The stability of rotating flows has been studied extensively since the early 20th century. These Taylor-Couette experiments were
carried out to study the fundamental
problem of turbulence \citep{Taylor23}. It had been recognised by \cite{Rayleigh17} that these rotating shear flows should be linearly unstable if the rotation profile between inner and outer cylinder 
possess a radially decreasing specific angular momentum and otherwise be stable. But accretion disks with their radially increasing specific angular momentum should be linearly stable. A non linear instability has so far not been found for the Keplerian flow \citep{Avila12}. But these experiments have been conducted using an incompressible working fluid, i.e.\ water, which means
no stratification has to be taken into account for stability.

This is fundementally different for gas disks around young stars.
In protoplanetary disks, large dust particles clearly want to orbit the star with Keplerian velocity. But the gas has typically a vertical and radial pressure gradient which alters its equilibrium 
rotation profile to a value slightly less than Keplerian velocity. This new velocity is a function of distance and introduces an additional vertical shear in the gas velocity, if one takes the radial temperature profile of the disk under consideration \cite{fromangetal11}. These modifications 
of the rotation profile have been extensively studied in the context of the interior of rotating stars. Therefor, it is not unreasonable to think of gas disks as very fast rotating stars but with much strong shear. 

This radial and vertical stratification, and thus the modification of the rotation profile, is highly important for several reasons. On one side, it makes the gas disk susceptible to hydro-dynamical instabilities
and on the other side, it complicates the dynamics of dust and ice grains. In disks that are sub-Keplerian, marginally coupled dust grains drift inward towards the star. But, they can be trapped in local laminar flow features like zonal flows and vortices. On top of that, a cloud of particles drifting inwards can produce the so called streaming instability, which is a candidate for an instability at work during the rapid formation of planetesimals.

\begin{figure}[ht]
	\centering
	\includegraphics[width=0.6\hsize]{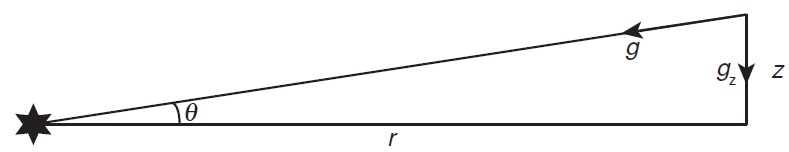}
	\caption{Cylindrical coordinate set up for a disk geometry (R,z), relating to the spherical coordinates (r,$\theta$), also indicating the cause for the vertical component of gravity in the disk. Only the cylindrical radial component of stellar gravity can be compensated by centrifugal acceleration. \citep{Armitage_2013}}
	\label{geometry}
\end{figure}

The vertical structure of a gas disk is given by the equilibrium of the vertical components of gravity, induced by the star, and by the pressure gradient of the gas.
The (spherical) radial component of gravity is $g = -\frac{G M}{r^2}$, which can be separated into a cylindrical radial component 
\begin{equation}
g_R = -\frac{G M}{r^2} \frac{R}{r}
\end{equation}
and a vertical component: 
\begin{equation}
g_z = -\frac{G M}{r^2} \frac{z}{r}.
\end{equation}
By definition $r^2 = R^2 + z^2$ and as long as the disk is thin ($z\ll R$) we can assume $R\approx r$. Thus, we receive the convenient equation:
\begin{equation}
g_z = -\frac{G M}{R^3} z = - \Omega^2 z,
\end{equation}
with $\Omega$ being the Keplerian orbital frequency in the mid-plane of the disk.

In the most simple case, the gas disk has no vertical temperature structure. In that case 
the gas pressure $P$ is a linear function of the density $P = c_s^2 \rho$, with $c_s$ being the
isothermal sound speed. The equations that we derive in the following are valid for arbitrary 
equations of state in which pressure is a potentially complicated function of density and temperature,
which can include phase transitions, i.e. variable mean molecular weight and changes in the degrees of freedom
of the gas molecules or atoms, thus changing the adiabatic index of the gas. As a result the true speed of
sound in the gas is given by $c_0^2 = \gamma \frac{\partial P}{\partial \rho}$, which only for $\gamma = 1$ would lead to the
isothermal speed of sound. Nevertheless, the definition of $c_s$ is practical for the structure equations of disk, despite the fact
that pressure fluctuations propagate at a slightly higher speed. 
Exploiting the Euler equation and using the definition of $c_s$ gives
\begin{equation}
0 = -\frac{1}{\rho} \nabla p - \Omega^2 z = -\frac{c_s^2}{\rho} \nabla \rho - \Omega^2 z,
\end{equation}
This differential equation is solved by a Gaussian density distribution
\begin{equation}
\rho(z) = \rho_0 e^{-\frac{z^2 \Omega^{2}}{2 c_s^2 }}= \rho_0 e^{-\frac{z^2}{2 H^2}},
\end{equation}
which contains the definition of the width of a Gaussian $H = \frac{c_s}{\Omega}$.
$H$ is usually called \textit{pressure scale height}, it is thus a parameter that describes the thickness of a disk, i.e. the
height at which the density, or pressure, has dropped to $60\%$ of the corresponding mid-plane value. 
Realistic disks are usually not strictly vertically isothermal. Yet, the temperature
variations are small in comparison to the density structure itself. Hence, the pressure scale height 
concept, based on the isothermal sound speed within the mid-plane, is still a good order of magnitude estimate for
the disk thickness and thus disk temperature.

\section{Protoplanetary Disk Evolution}
\begin{figure}[ht]
	\centering
	\includegraphics[width=0.6\hsize]{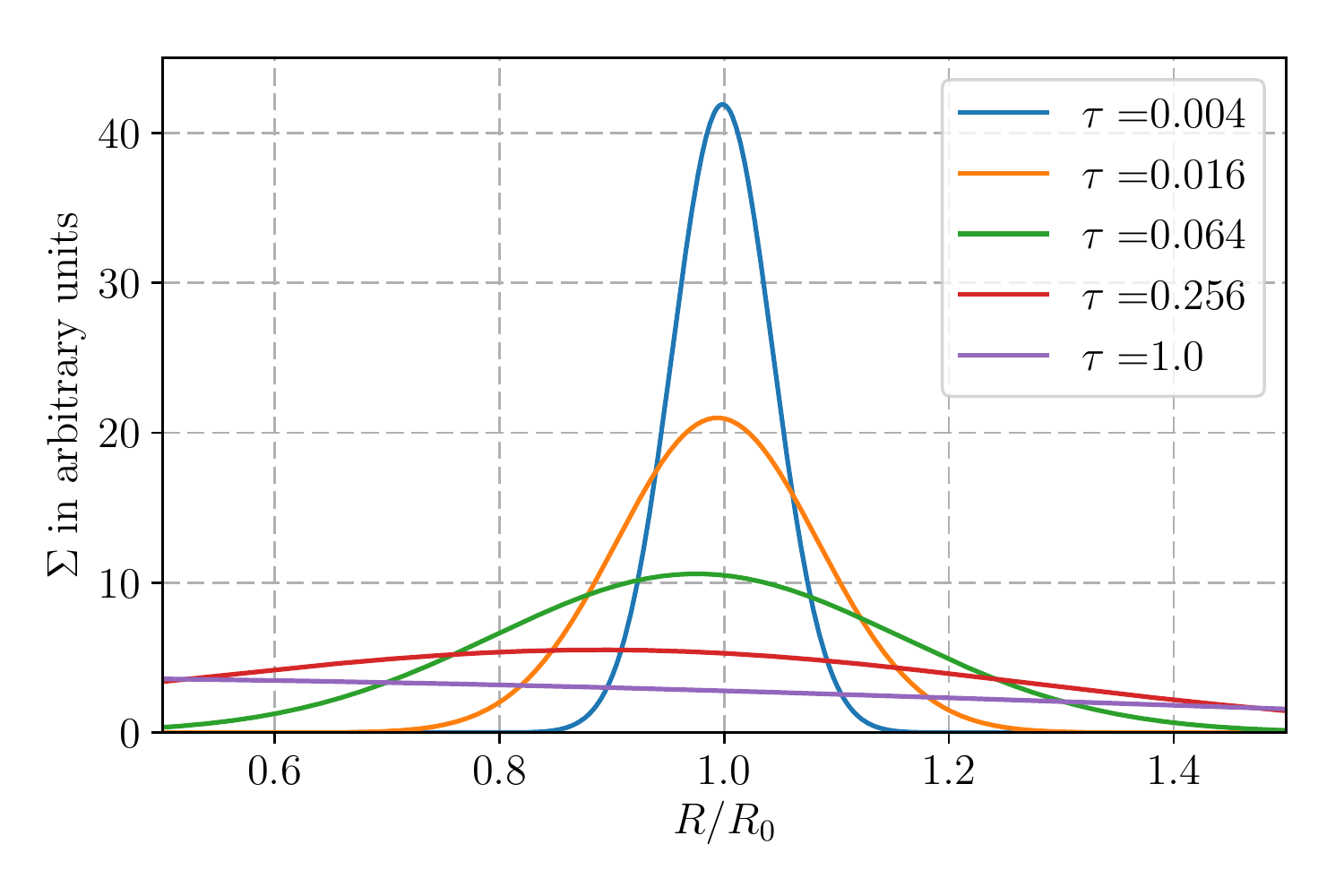}
	\caption{Solution of the disk evolution equation for different times $\tau$ plotting the results from \citep{Pringle_1981}.}
	\label{evolution}
\end{figure}

Protoplanetary disks are not static structures. This conclusion can be drawn because of observations of the spectral energy distribution of young stellar objects. Due to the amount of warm dust inside of the protoplanetary disks, their spectra show an IR-excess. Another characteristic is an UV-excess, caused by hotspots on the stellar surface, which are produced by infalling material from the disk onto the star. By determining the time at which the IR or the UV-excesses of young stellar objects disappear, one can estimate the timescale at which almost all of their disk material has been accreted. \cite{Haisch_2001} and \cite{2014prpl.conf..475A} have determined a lifetime of $10^6-10^7$ years by means of this method. This found timescale equals thousands of dynamical timescales ($\tau_{dyn}=\Omega^{-1})$, leading to the assumption that protoplanetary disks are slowly evolving system with quasi-static structures. 
General models of the evolution of the mass distribution of geometrically thin disks have been developed by \cite{Luest_1952} and \cite{Pringle_1981} and require some form of viscous dissipation, leading to the outward transport of angular momentum. Material with decreasing kinetic energy hence moves inwards. This process is what we define as accretion. 
In the case of a geometrically thin disk, it can be described with the disk evolution equation, by \cite{Pringle_1981}:
\begin{equation}\label{diskev}
	\frac{\partial \Sigma}{\partial t}=\frac{3}{R}\frac{\partial}{\partial R}\left[ R^{\frac{1}{2}}\frac{\partial}{\partial R}(\nu \Sigma 	R^{\frac{1}{2}})\right]
\end{equation} 
Without specifying the physics leading to the kinematic viscosity $\nu$, equation \eqref{diskev} can give a first impression on how viscosity acts on a protoplanetary disk. 
This can be visualised by the solution for an initially $\delta$-distributed surface density.
The following equation was derived by \cite{Pringle_1981} and shows, that most of the material is moving inwards, while a small portion of it carries the angular momentum away (Figure \ref{evolution}): 
\begin{align}
	\Sigma(R,\tau=0)&=\frac{m}{2\pi R_0}\delta(R-R_0) \\
	\Sigma(x,\tau)&=\frac{m}{\pi R_0^2 \tau x^{\frac{1}{4}}}exp\left[\frac{-(1+x^2)}{\tau}\right]I_{\frac{1}{4}}\left(\frac{2x}{\tau}\right)
\end{align}	
$I_{\frac{1}{4}}$ is a modified Bessel-function of the first kind, $x=\frac{R}{R_0}$ and $m$ represents the total mass of the distribution.
The separation of mass and angular momentum, as a consequence of viscous friction, is also a possible explanation for the solar systems angular momentum problem. (The planets carry just $\sim 1\%$ of the systems mass, while having $\sim 99\%$ of its angular momentum) \cite[p. ~3]{Armitage_2013}.
Equation \eqref{diskev} has the form of a diffusion equation and thus defines a theoretical viscous-evolution timescale:
\begin{equation}\label{tdyn}
	\tau_{ev}\approx\frac{R^2}{\nu}
\end{equation}
\cite{Armitage_2013} calculated this timescale under the assumption that the internal dissipation of kinetic energy is driven by molecular collisions (molecular viscosity).
The result is an evolution time of $\tau_{ev}\approx10^{13}$ years, which is seven orders of magnitude longer than the observed evolution times, derived from the spectra of young-stellar-objects and three orders of magnitude larger than the age of the observable universe.
This means, that molecular viscosity can not be the only reason for angular momentum transport in protoplanetary disks.
\cite{Shakura_1973} quantified the effective viscosity, caused by turbulent friction with the nowadays widely used $\alpha$-description, as defined above.
Several physical processes can lead to turbulent motion within a protoplanetary disk and therefore drive angular momentum transport. In the following we will have a look on the most prominent ones and will come to the conclusion that a laminar disk is highly unlikely.

\subsection{Magnetorotational Instability}

One of the most promising sources is the \textit{magnetorotational instability} (MRI) \citep{Balbus_Hawley_1991}. It requires the presence of a weak magnetic field and a partially ionised gaseous shear flow. Due to the ionisation, the gas particles are coupled to the magnetic field which also can be distorted by the particles movement. 
The underlying mechanism, that makes this configuration linearly unstable is the magnetic tension coupling of fluid parcels on different orbital radii. In a simple picture one can consider two charges fluid elements on different orbital radii, which are coupled that way. If the angular velocity profile follows the Keplerian angular velocity, the fluid parcel on the outer orbit drifts away from the inner one. The field line connecting the fluid elements causes magnetic tension forces which transfer angular momentum from the inner to outer elements acting, in effect,  as a repelling force.
The consequence is an accelerated drift away and thus an unstable flow.
This mechanism can provide an effective $\alpha$-value of $\alpha\approx10^{-2}$ in simulations under idealistic conditions \citep{Davis_2010} for perfect ionisation. 

But, for a realistic setup of a disk, still today it is not known how strong  $\alpha$ can be in a simulation of the MRI a stratified disk. Numerical simulations have not converged, and increasing resolution seems to lower the numerical determined $\alpha$-values \citep{2017ApJ...840....6R}.
 
Additionally, protoplanetary disks are mostly weakly ionised structures and non-ideal magneto-hydro-dynamical effects can further damp the evolution of the MRI, producing a dead zone in the disks dense interior. \cite{Dzyurkevich_2013} mapped the extent of this dead zone arising from low number of electrons, e.g. resistivity, in simulations and found the dead zone to have widths of up to $12 AU$ in the radial direction.
This leads to the question, whether purely hydro-dynamical instabilities can also drive angular momentum transport in the MRI-inactive parts of protoplanetary disks.

Recently additionally ambipolar diffusion has been identified to damp the MRI in the upper atmospheres of disks as well as at large distances from the star, because at low gas densities the collisions of ions and neutrals become to rare for an effective coupling to the magnetic field \citep{2013ApJ...769...76B}, yet still allowing for accretion via launching a wind \citep{2017A&A...600A..75B}.

The latest development in the field of the MRI involves the inclusion of the non-ideal Hall term, which does not create additional turbulence, but leads to the generation of mean magnetic fields in the dead zone and interesting flow features, like vortices and zonal flows \citep{2014A&A...566A..56L,2016A&A...589A..87B}.

\section{Hydrodynamic Stability}\label{hydrostab}
On the other hand, if non-ideal MHD effect largely hamper the MRI, they open a venue for other hydro-dynamic instabilities,
as suggested by \cite{2011A&A...527A.138L}.
Before going into the individual possible instabilities, we present the derivation of the Rayleigh criterion
as the only explicit derivation in this text. This is because the linear stability of the Keplerian flow as
studied in Taylor Couette flows in the laboratory is of such a fundamental importance for disks \citep{2006Natur.444..343J}. It is the most simple criterion as it does not involve any self-gravity, stratification in pressure and entropy, thermal relaxation, nor magnetic fields, yet all stability analyses involving these effects are variations of this fundamental stability investigation.

\subsection{Rayleigh-Criterion}
In order to investigate the general hydro-dynamic stability of a protoplanetary disk, one can approximate the gas flux as a circular shear flow within two rigid cylinders (Taylor-Couette-flow). Whereas a stratified shear flow can be linearly unstable (Kelvin Helmholtz Instability), the plane Couette flow is only non linear unstable. So if the Keplerian flow would be linear unstable because of the underlying shear, then the problem of turbulence in disks would have an easy solution.

To show that a pure unstratified Keplerian flow is unfortunately stable, a stability analysis can be done, for example following \cite[p.~71]{Drazin_Reid_2004} or \cite[p.~158]{Pringle_King_2007}. The equations of fluid motion are therefore written in cylindrical coordinates as follows:
\begin{eqnarray}
\label{fluideq}
	\frac{D\vec{v}}{Dt}=\frac{\partial \vec{v}}{\partial t}+(\vec{v}\cdot \vec{\nabla})\vec{v}&=-										\frac{\vec{\nabla}p}{\rho} \\
	\frac{\partial v_R}{\partial t}+v_R\frac{\partial v_R}{\partial R}+\frac{v_{\phi}}{R}\frac{\partial v_R}{\partial \phi}-			\frac{v_{\phi}^2}{R}+v_z\frac{\partial v_R}{\partial z}&=-\frac{1}{\rho}\frac{\partial p}{\partial R}\nonumber\\
	\frac{\partial v_{\phi}}{\partial t}+v_R\frac{\partial v_{\phi}}{\partial R}+\frac{v_{\phi}}{R}\frac{\partial v_{\phi}}				{\partial \phi}+\frac{v_{\phi}v_R}{R}+v_z\frac{\partial v_{\phi}}{\partial z}&=-\frac{1}{R\rho}\frac{\partial p}{\partial \phi}\nonumber\\
	\frac{\partial v_z}{\partial t}+v_R\frac{\partial v_z}{\partial R}+\frac{v_{\phi}}{R}\frac{\partial v_z}{\partial \phi}+v_z			\frac{\partial v_z}{\partial z}&=-\frac{1}{\rho}\frac{\partial p}{\partial z}\nonumber
\end{eqnarray}
The continuity equation in cylindrical coordinates is:
\begin{eqnarray}
\label{continuity}
	\frac{\partial \rho}{\partial t}+\vec{\nabla}\cdot(\rho\vec{v})&=0\\
	\frac{\partial \rho}{\partial t}+\frac{1}{R}\frac{\partial (R\rho v_R)}{\partial R}+\frac{1}{R}\frac{\partial \rho 					v_{\phi}}{\partial \phi}+\frac{\partial \rho v_z}{\partial z} &=0\nonumber 
\end{eqnarray}
To do the stability analysis, these equations have to be simplified by adding a velocity perturbation field $\vec{v}'=(v_r',v_{phi}',v_z')$ and defining the initial unperturbed flow. 
Therefore, the following assumption are made:
\begin{itemize}
	\item
	The unperturbed velocity field is given by $\vec{v}_0=(0,V(R),0)$ and related to the angular frequency via $V(R)=\Omega(R)R$ (Taylor-Couette-flow).  
	\item 
	The fluid is incompressible, so that $\rho=constant$. This leads to the incompressible continuity equation $div(\vec{v})=0$, which is already linearised.
	\item
	Since $\rho$ is constant, the pressure is given by the centrifugal forces acting on a fluid parcel: $p(R)=\rho\int\frac{V(R)^2}{R}dR$
	\item
	For the linear analysis, all terms containing products of perturbations are neglected. Perturbations are assumed to be infinitesimal and indicated by primes. 
\end{itemize}
This concludes into a new set of, now linearised, equations:
\begin{eqnarray}
\frac{\partial v_R'}{\partial t}+\Omega (R)\frac{\partial v_R'}{\partial \phi}-2\Omega (R) v_{\phi}'&=-\frac{1}{\rho}\frac{\partial p'}{\partial R}\\
\frac{\partial v_{\phi}'}{\partial t}+v_R'\left(\frac{d V(R)}{d R}+\Omega (R)\right)+\Omega (R) \frac{\partial v_{\phi}'}{\partial \phi}&=-\frac{1}{\rho R}\frac{\partial p'}{\partial \phi}\\
\frac{\partial v_z'}{\partial t}+\Omega(R)\frac{\partial v_z'}{\partial \phi}&=-\frac{1}{\rho}\frac{\partial p'}{\partial z} \\
\frac{\partial v_R'}{\partial R}+\frac{v_R'}{R}+\frac{1}{R}\frac{\partial v_{\phi}'}{\partial \phi}+\frac{\partial v_z'}{\partial z}&=0
\end{eqnarray}
The next step is to consider all perturbed quantities (including the pressure $p \rightarrow p'$) to have a wavelike form e.g. $v_R(R,\phi,z)'=v_R(R)'exp[i(\omega t+m\phi+kz)]$ with $m \in \mathbb{Z}$. This step is done to Fourier transform the equations of motion in order to analyse the spectrum of perturbation wavelengths. This results in three further simplified equations, describing the evolution of wave-like perturbations:
\begin{eqnarray}
	i(\omega+m\Omega(R))v_R'-2\Omega(R) v_{\phi}&=-\frac{1}{\rho}\frac{d p'}{d R}\label{linearised1}\\
	i(\omega+m\Omega(R))v_{\phi}'-v_R'\left(\Omega(R)+\frac{d V(R)}{d R}\right)&=-\frac{im p'}{R\rho}\label{linearised2}\\
	i(\omega+m\Omega(R))v_z'&=-ik\frac{p'}{\rho} \\
\frac{d v_R'}{dR}+\frac{v_R'}{R}+\frac{i m v_{\phi}'}{R}+ikv_z'&=0	
	\label{linearised3}
\end{eqnarray}
These expressions are written from an Eulerian point of view, which means that they are describing a specific, fixed point in space and the time evolution of the flow in that position. Therefore, the linearised set of equations \eqref{linearised1} to \eqref{linearised3} can not state how a specific fluid element behaves when its trajectory gets perturbed. In this fashion, it is convenient to describe the displacement of a fluid parcel in a Lagrangian way, which means to follow a perturbed fluid element along its path. The set of linearised equations can be rewritten by expressing the velocity perturbation field $\vec{v}'$ via the Lagrangian spatial displacement $\vec{\xi}=\vec{r}-\vec{r}_0$, where $\vec{r}_0$ defines the unperturbed and $\vec{r}$ the perturbed trajectory of a fluid element. The velocity is therefore defined via:
\begin{equation}
	\vec{v}'=\frac{\partial \vec{\xi}}{\partial t} +(\vec{v}_0\cdot \vec{\nabla})\vec{\xi}-(\vec{\xi}\cdot								\vec{\nabla})\vec{v}_0
\end{equation}
\cite{Drazin_Reid_2004} obtained a final equation for the perturbed fluid flow, by using these relations and assuming only axisymmetric perturbations ($m=0$):
\begin{eqnarray}\label{lasteq}
	\frac{d}{dR}\left(\frac{1}{R}\frac{d}{dR}(R\xi_R)\right)-k^2\xi_R=-\frac{k^2 \mathcal{R}(R)}{\omega^2}\xi_R
\end{eqnarray}
where $\mathcal{R}(R)$ is the Rayleigh discriminant, also known as the square of the epicyclic frequency $\kappa$. 
Solving this equation for $\omega$ gives the Rayleigh criterion for the stability of a Taylor-Couette-flow under an axisymmetric perturbation. 
The motion is stable, if and only if perturbing it causes epicyclic oscillations around the equilibrium state. This means that $\kappa^2$ has to be a positive number.
If $\kappa^2$ is negative, the perturbation $\xi_R$ will grow exponentially, causing an instability.
Therefore, the Rayleigh criterion follows from equation \eqref{lasteq}:
\begin{equation}\label{Rayleigh}
	\kappa^2(R)=\mathcal{R}(R)=\frac{2\Omega (R)}{R}\frac{d(R^2\Omega (R))}{dR} >0 \qquad \textit{Stability}
\end{equation}
The gas of a protoplanetary disk orbits the star with an angular frequency close to $\Omega=\sqrt{\frac{GM_*}{R^3}}$, which leads to the conclusion that protoplanetary disks are generally stable flows in the sense of the Rayleigh criterion.

\subsection{Toomre-Criterion}\label{Toomre}
One special instability case occurs at the time the disk is forming, e.g. when more mass is falling onto the disk than can be transported away: then the disk can become very massive and self-gravity will drive the formation of spiral arms and angular momentum transport.

The previous investigations of a disks stability did not consider the self-gravity of the gas. To do so, the following section sums up the derivation of the Toomre criterion by \cite[p.~172]{Pringle_King_2007}. They approximated the disk as a rotating sheet with mass distribution $\rho(R,z)=\Sigma(R)\delta(z)$. 
The unperturbed velocity field is defined as in the previous sections. It is convenient for this derivation to assume a constant pressure and density profile. Therefore $\vec{\nabla}p$ vanishes and the hydrostatic equilibrium just states:
\begin{equation}\label{toomreequilibrium}
\Omega^2=-\frac{1}{R}\frac{\partial \Phi}{\partial R}.
\end{equation} 
$\Phi$ denotes the unperturbed gravitational potential. Thus, equation \eqref{toomreequilibrium} describes the balance of gravitational and centrifugal force.
The perturbed velocity profile is given by $\vec{v}'=(v_R',R\Omega+v_{\phi}')$. For simplicity, all perturbations are assumed to have the form $v_R'=v_R(R)\exp(i\omega t)$.
This leads to a set of two linearised momentum equations \eqref{momentum1} and \eqref{momentum2}, and the continuity equation \eqref{cont}:
\begin{eqnarray}
i\omega v_R'-2\Omega v_{\phi}'&=-\frac{1}{\Sigma}\frac{d p'}{dR}-\frac{d\Phi'}{dR} \label{momentum1}
\\
i\omega v_{\phi}'+\left(\Omega+\frac{d(R\Omega)}{dR}\right)v_R' &=0 \label{momentum2}\\
i\omega \Sigma'+\Sigma\left(\frac{dv_R'}{dR}+\frac{v_R'}{R} \right)&=0 \label{cont}
\end{eqnarray}  
The perturbed gravitational potential is expressed in terms of a short wavelength perturbation by using Poisson's equation.
Derivatives are therefore replaced by simply multiplying with $ik$, due to the Fourier-ansatz. 
This leads to:
\begin{equation}
\Phi'(z=0)=-\frac{2\pi G \Sigma'}{|k|}
\end{equation}
By combining the perturbed equations and the gravitational potential, an equation can be found that relates $\omega$ with $k$ (dispersion relation):
\begin{equation}
\omega^2=\kappa^2-2\pi G |k|\Sigma+k^2 C_s^2,
\end{equation}
with the two-dimenional speed of sound $C_s^2=\frac{dP}{d\Sigma}$. This expression becomes the rotation modified dispersion relation  for sound waves ($\omega^2 =C_s^2 k^2+\kappa^2$), if self-gravity is negligible. Instability will occur if $\omega^2<0$ holds. Therefore, the Toomre criterion for gravitational instability follows:
\begin{equation}\label{Toomre_Crit}
Q=\frac{\kappa C_s}{\pi G \Sigma}<1 \qquad \textit{Instability}
\end{equation}
It defines the relation of destabilising gravitational and stabilising centrifugal and pressure forces under which the disk is unstable due to its own gravity. 

Self-gravity will therefor lead to a very active turbulence, angular momentum transport and mass accretion onto the star. Yet, this phase is short-lived. After the mass infall onto the disk stops, the disk mass will fall below the conditions for Toomre instability and render a purely pressure supported, almost Keplerian disk and the stability is governed by the Solberg-H\o iland Criteria.

\subsection{Solberg-H\o iland Criteria}
Disks have both a radial and vertical stratification in density and temperature, determined by heating processes like viscosity or irradiation and by cooling via radiation. This stratification will be hydro-static for the equilibrium state, but the question is what happens to perturbations in entropy? Will a gas parcel be stabilised by buoyancy or drive a convective instability?

The stability to convection can be expressed by the buoyancy-frequency $N^2$ (aka Brunt-V\"{a}is\"{a}l\"{a} frequency), which can be determined for the radial and for the vertical stratification as $N_R^2$ and $N_z^2$.

An intuitive approach was presented by \cite{Shore_2007}. If a fluid element is displaced adiabatically upwards, due to a perturbation, it will be hotter than its surroundings (in the case of an internally heated disk). Consequently, it will experience a buoyant force, that will cause a further upwards acceleration.
This accelerated rise will happen if the disks temperature gradient becomes steeper than adiabatic. Schwarzschild's criterion for convective stability can therefore be reformulated as:
\begin{equation}\label{superad}
\left|\left(1- \frac{1}{\gamma}\right)\frac{T}{p}\vec{\nabla}p  \right|=\left|\vec{\nabla}T\right|_{ad}>\left|\vec{\nabla}T\right| \qquad \textit{Stability}
\end{equation} 
Stability is thus characterised by a real valued buoyant frequency, at which a fluid parcel oscillates around its equilibrium position in a stable stratified fluid \cite{Ruediger_2002}:
\begin{align}
N_R^2&= - \frac{1}{\gamma \rho} \frac{\partial p}{\partial R} \frac{\partial}{\partial R} log\left(\frac{p}{\rho^{\gamma}}\right)
\\
N_z^2&= - \frac{1}{\gamma \rho} \frac{\partial p}{\partial z} \frac{\partial}{\partial z} log\left(\frac{p}{\rho^{\gamma}}\right) 
\end{align}

\cite{Ruediger_2002} derived the Solberg-H\o iland conditions for stability in the presence of rotation in accretion disks. They performed a complete linear stability analysis of the hydro-dynamical set of equations in cylindrical coordinates. 
The resulting criteria incorporate the stabilising effect of rotation, that was not included in the standard Schwarzschild-Criterion:
\begin{eqnarray}
	N_R^2+N_z^2+\kappa^2&>0 \label{Solberg1}\qquad \textit{Stability}\\
	\frac{\partial p}{\partial z} \left( \kappa^2 \frac{\partial}{\partial z} log \left( \frac{p}{\rho^{\gamma}} \right) - R 			\frac{\partial \Omega^2}{\partial z} \frac{\partial}{\partial R} log\left(\frac{p}{\rho^{\gamma}_0}\right)\right)&<0 \label{Solberg2}\qquad 		\textit{Stability},
\end{eqnarray}
where $\kappa$ is the epicyclic frequency, which is equal to the Keplerian angular frequency $\Omega$ in case of Keplerian rotation.
It can now be shown that by introducing thermal relaxation into this system, then there can be unstable, growing modes.

\section{Linear Hydrodynamic Instabilities}\label{hydroinstab}
Despite the fact that their angular momentum distribution makes protoplanetary disks linearly stable in the sense of the Rayleigh criterion, they can develop local hydro-dynamic instabilities if the Solberg-H\o iland criteria is violated. The non-trivial temperature stratification in a disk produces misaligned isobars and isopycnals (surfaces with constant density), leading to vertical shear i.e. a vertical gradient in the rotation profile. This state is called baroclinic and also exists in planetary atmospheres, where it causes large scale weather patterns like cyclones and anticyclones on earth or Jupiter's great red spot. Rotating stars also exhibit a baroclinic stratification, which leads to the Goldreich-Schubert-Fricke Instability \citep{Goldreich_Schubert_1967, Fricke_1968}. In the case of a protoplanetary disk, a baroclinic stratification can become unstable if the disk is vertically buoyantly neutral or radially buoyant and if radiative cooling of perturbations is taken into account. The Solberg-H\o iland criteria therefore have to be modified for thermal relaxation.
In the past decades, attempts have been made to explain if and how purely hydro-dynamical instabilities can lead to the dynamical evolution of the MRI-inactive dead zones.

The instabilities  found so far arise either from vertical shear or from the radial buoyancy in the disk. We will see that the \textit{Vertical Shear Instability} is a violation of Rayleigh Criterion and that \textit{Convective Overstability} and its weakly non-linear extension Subcritical Baroclinic Instability are both special cases of thermal convection in the radial direction of the disk. The following sections briefly review these mechanisms.

\subsection{Vertical Shear Instability}\label{subsec:vsi}
The Vertical Shear Instability (VSI) is the analogue of the well studied Goldreich-Schubert-Fricke Instability (GSF) \citep{1967ApJ...150..571G, 1968ZA.....68..317F} for protoplanetary disks. This instability was initially discussed in the context of rotating stars, where it is caused by a baroclinic stratification. The VSI can develop if the disk's angular frequency has a vertical gradient. Vertically perturbed fluid parcels, which move along the iso-surfaces of angular momentum, gain kinetic energy and circumvent Rayleigh's stability criterion \citep{1998MNRAS.294..399U} (See Fig.~\ref{VSI_crit} for a schematic explanation). This instability therefore drives modes, which are vertically elongated ($k_R/k_z \gg 1$) \citep{Urpin_2004}.
\citet{Nelson_2012}, who were the first to show that the VSI can operate in protoplanetary disks, calculated the corresponding growth rate for a locally isothermal, compressible gas under the shearing sheet approximation  
\begin{equation}\label{VSI_growth}
\Gamma_{\mathrm{VSI}}^2=\frac{-\kappa_R^2(c_s^2 k_z^2+N_z^2)+2\Omega c_s^2 k_R k_z (\partial_z v_{\phi}(R,z))}{c_s^2(k_z^2+k_R^2)+\kappa_R^2+N_z^2},
\end{equation}
where $N_z$ is the vertical buoyancy frequency, $c_s$ is the local sound speed, $v_{\phi}$ is the azimuthal velocity, and $\Omega$ is the Keplerian angular frequency. They performed numerical simulation of the instability and found narrow, almost vertical motions, which caused $\alpha \sim 10^{-3}$.

\begin{figure}[ht]
\centering
\includegraphics[width=1.0\textwidth]{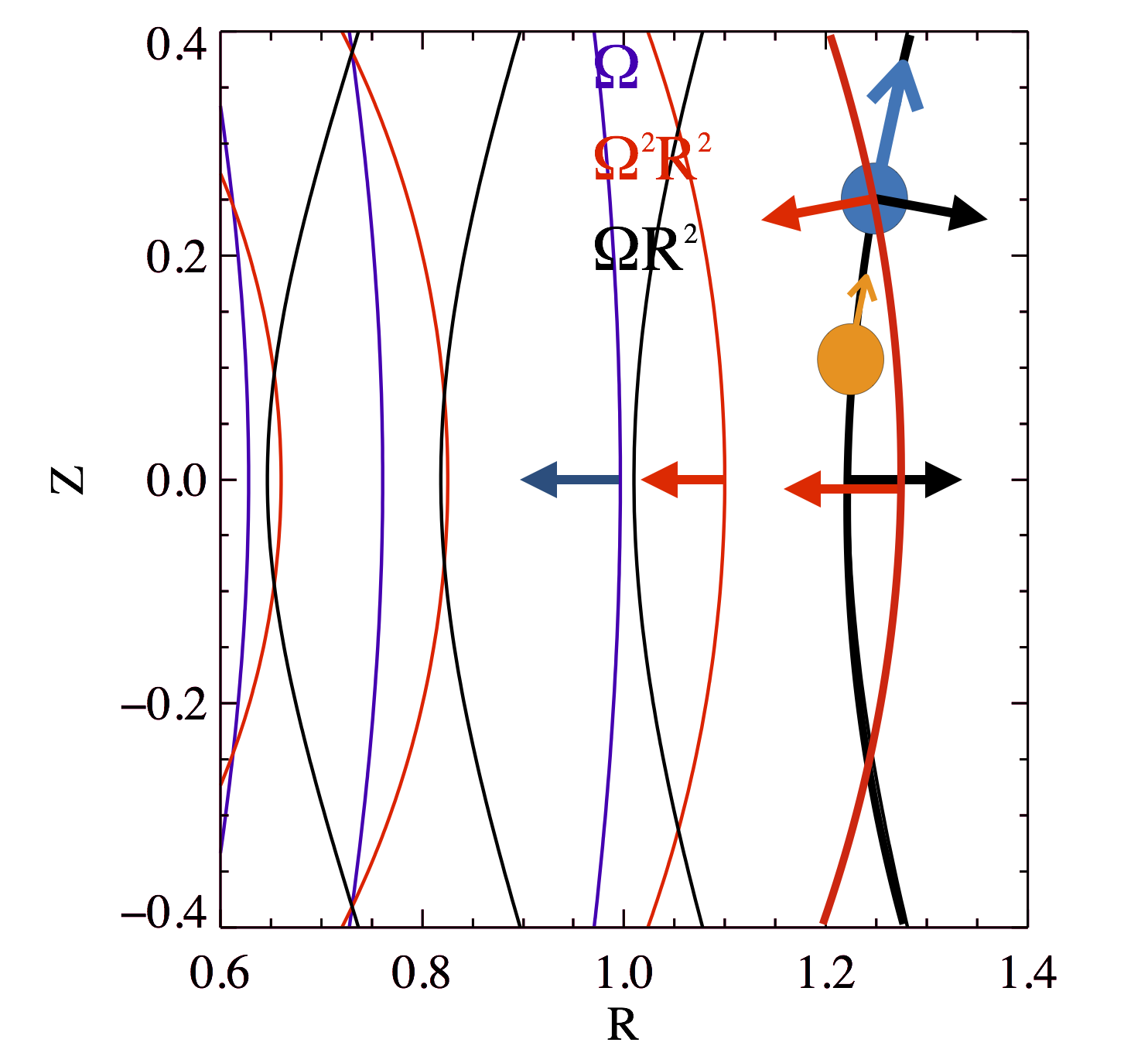}
\caption{Schematic Vertical Shear Instability mechanism: In disks with a radial temperature gradient gas does not rotate on cylinders, but the contours of constant angular frequency $\Omega(R,z)$ (blue contours) are bent inward. While $\Omega \propto R^{-1.5}$ and the associated kinetic energy $\Omega^2 R^2 \propto R^{-1}$ (red contours) decrease with radius (see also the blue and red arrow, indicating the gradient towards larger values), the specific angular momentum increases  as $\Omega R^2 \propto R^{0.5}$ (black contours and outward pointing arrow). This means that a gas parcel (yellow circle) if slightly pushed upward, can continue to move upward-outward under conservation of angular momentum (not being pushed back, which would for instance happen if kicked strictly radially, leading to epicyclic oscillations). As at the same time the gas parcel is carrying more kinetic energy than the surrounding gas, it continues to accelerate, as indicated by the blue circle. This is the same process as in a Rayleigh unstable disk without vertical shear, but where the radial specific angular momentum radially rises $\Omega(R) \propto R^{-2}$. Note that vertical motions can only be excited if the disk is vertically adiabatic or fast thermal relaxation neutralizes a vertically stable stratification in entropy. Design by the authors.}
\label{VSImech}
\end{figure}

In this work, we are interested in the maximal growing perturbation of a certain radial wavenumber. Therefore, we use the following condition by \citet{Urpin_2004} to get the corresponding fastest growing vertical wavenumber
\begin{equation}\label{VSI_max}
k_z=\frac{k_R
}{2}\frac{\partial_z j^2(R,z)}{\partial_R j^2(R,z)},
\end{equation}
where $j(R,z)$ represents the specific angular momentum.
The vertically perturbed fluid parcel's which are prone to be unstable to the VSI experience buoyant forces in the case of a stable stratification, which impede the instability's growth. Fast thermal relaxation can overcome this effect, because it adjusts the fluid parcels temperature to the background temperature and therefore diminishes buoyancy driving entropy differences.
\citet{Lin_2015} considered this effect and derived a critically slow relaxation time scale for which the VSI can grow
\begin{equation}
\tau_{\mathrm{crit}}\lesssim \frac{\left|\partial_z v_{\phi}\right|}{N_z^2}\label{VSI_crit}.
\end{equation}
A convectively unstable or neutral vertical stratification does not impede vertical perturbations and thus allows for VSI in the presence of sufficiently fast cooling. The behavior of VSI in such a polytropically stratified disk was also  studied by \cite{Nelson_2012}, who found a critical cooling time of $\tau_{\mathrm{relax}}\Omega \approx 10$. It is therefore necessary to investigate whether the stratification is potentially unstable to convection, since the resulting change in disk structure might allow for VSI, even if criterion \eqref{VSI_crit} is not fulfilled.

\subsection{Convective Overstability}\label{subsec:cos}
\begin{figure}[ht]
\centering
\includegraphics[width=1.0\textwidth]{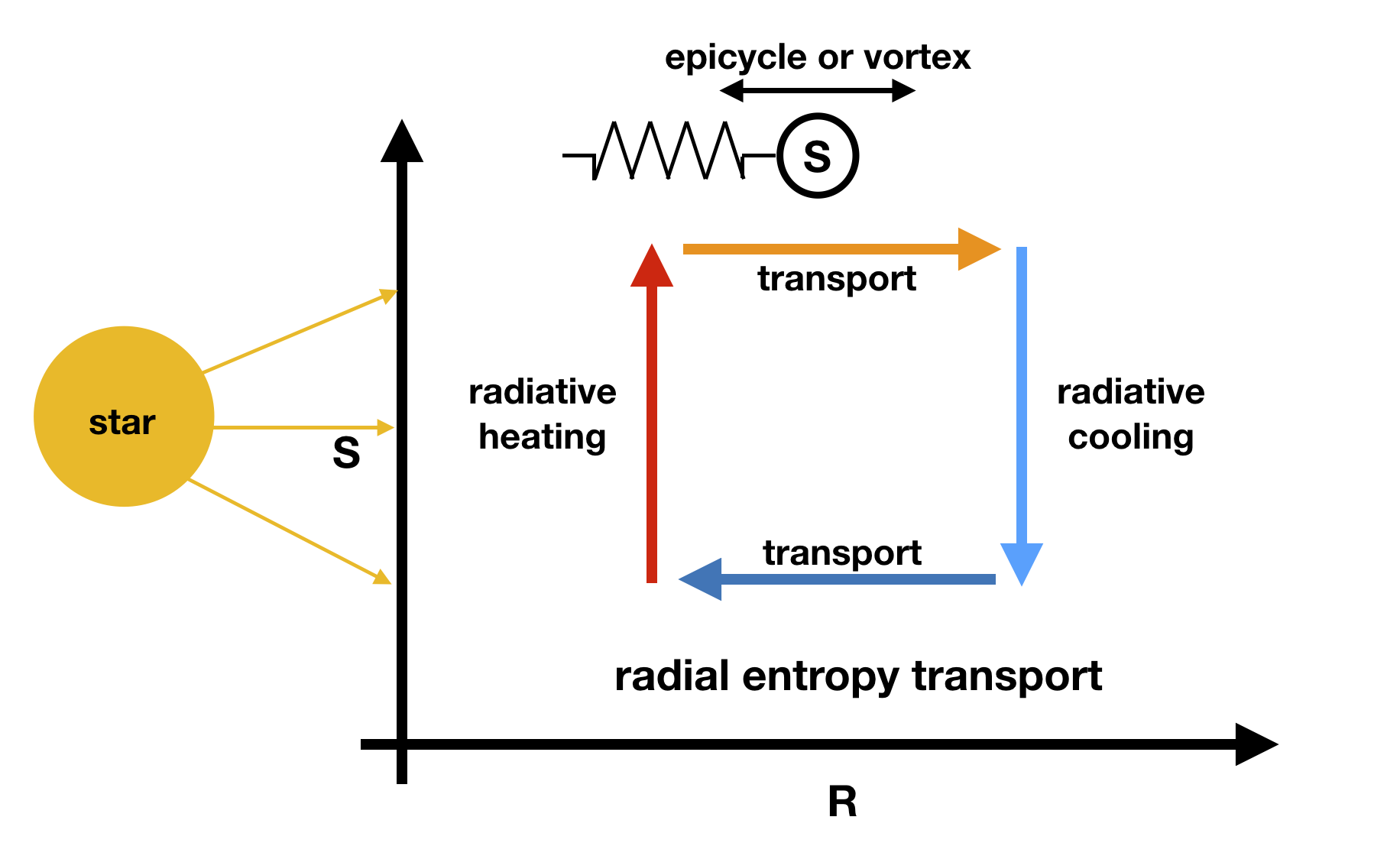}
\caption{Schematic Convective Overstability and Subcritical Baroclinic Instability mechanism. The epicyclic oscillation of a radially perturbed fluid element carrying entropy $s$ is shown. During radial inward-outward motion through the disk, it radiatively thermalises with its surrounding on the local relaxation time scale, which leads to buoyancy forces, enhancing its motion, while transporting entropy radially outward. SBI and COS modes are both in first order epicyclic motions, e.g. enforced via the rotation of the disk, with the COS modes small scale $r-z$-plane modes of frequency $\Omega$ and the SBI vortices large scale $r-phi$-plane modes that due to their internal pressure maximum can have rotation frequencies smaller than the epicyclic frequency $\omega < \Omega$. Design by the authors.}
	\label{COSmech}
\end{figure}

\citet{Klahr_Hubbard_2014} first considered finite thermal relaxation times ($\tau_{\mathrm{relax}}$) in their linear, inelastic stability analysis of protoplanetary disks and found a new, thermally driven instability, which can be described as radial convection on epicycles. The mechanism, schematically shown in \autoref{COSmech}, operates within disks that exhibits a radially buoyant stratification, e.g. entropy falls with radius. The outward displacement of a fluid element brings it into contact with a cooler surrounding. While the gas parcel undergoes half an epicycle, it changes its temperature on the local thermal relaxation time scale. When it arrives at the initial position, it is cooler than its surrounding. Additionally, the entropy decreases below the initial value and the fluid element experiences a buoyancy force, accelerating it inwards. On its half epicycle through the inner and hotter region of the disk, it undergoes an increase of temperature and entropy. When it finally arrives at the initial radius, it experiences an outward buoyancy force. A runaway process begins, leading to growing amplitudes. This positive feedback is called Convective Overstability (COS).

The linear phase of the COS drives motions in the disk's $R-z$ plane with a growth rate of  
\begin{equation}\label{COS_rate}
\Gamma_{\mathrm{COS}}=\frac{1}{2}\frac{-\gamma \tau_{\mathrm{relax}} N_R^2}{1+\gamma^2 \tau_{\mathrm{relax}}^2(\kappa_R^2+N_R^2)}
\end{equation}
\citep{Klahr_Hubbard_2014}, where $\gamma$ is the heat capacity ratio of the gas, $\kappa_R$ is the radial epicyclic frequency and $N_R$ refers to the radial Brunt-V\"ais\"al\"a-frequency.
Since this overstability relies on entropy differences between perturbed fluid parcels and their surrounding, relaxation times have to be neither too small nor too large. In the first case, a fluid element would always adopt the temperature of its surrounding. Its movement would be isothermal and no buoyant force would act on it, which means that Rayleigh's stability criterion applies to it. The latter case describes an adiabatic perturbation, where the fluid's entropy stays constant during its epicyclic motion. This means that it follows a stable, buoyancy adjusted epicycle \citep{Latter_2016}. 
The relaxation time for maximum growth of the linear phase was also calculated by \cite{Klahr_Hubbard_2014} to be
\begin{equation}\label{COS_max}
\tau_{\mathrm{max, COS}} =\frac{1}{\gamma \Omega},
\end{equation} 
where $\Omega$ is the local Keplerian angular frequency.

\subsection{Subcritical Baroclinic Instability}\label{subsec:sbi}
The finite amplitude perturbations created by the COS can trigger the Subcritical Baroclinic Instability (SBI) \citep{Klahr_Bodenheimer_2003, Petersen1_2007, Petersen2_2007}, which forms and amplifies vortices in the disk's $R-\phi$ plane. \cite{Lyra_2014} has explicitly shown in numerical simulations that that this instability is the saturated form the COS.
The vortices, created and amplified by this instability are of interest for the growth of dust to planetesimals, because they are able to accumulate dust particles \citep{Barge_1995}.
They have vertically little variation over more than a pressure scale height of the disk \citep{Meheut_2012, Manger_2018}. Thus to study their possible amplification in the SBI mechanism, which relies on the radial buoyancy, one has to consider their vertically integrated entropy and pressure structure, which means to treat them as quasi two-dimensional structures.
We therefore use the definition of a vertically integrated density and entropy, also used in \cite{Klahr_2004} and \cite{Klahr_2013}
\begin{align}
\Sigma &=\int_0^{z_{\mathrm{max}}} \rho(R,z)\mathrm{d}z \\
\tilde{S}&=C_{\mathrm{V}} \log(\tilde{K}),
\end{align}
where we assume a polytropic equation of state for a two dimensional pressure $\tilde{P} \propto \tilde{K}\Sigma^{\tilde{\gamma}}$ with an entropy-like potential temperature $\tilde{K}=\tilde{P}\Sigma^{-\tilde{\gamma}}$. The heat capacity ratio has to be adjusted to a vertically integrated value $\tilde{\gamma}$, defined by \cite{Goldreich_1986} as $\tilde{\gamma}=(3\gamma -1)/(\gamma+1)=1.354$ for a typical mixture of hydrogen and helium gas ($\gamma=1.43$). We now adopt the definitions of the logarithmic gradients in column density, vertically integrated pressure and entropy from \cite{Klahr_2004}, given by
\begin{align}
\beta_{\Sigma}&=\frac{\mathrm{d}\log (\Sigma)}{\mathrm{d} \log (R)} \\
\beta_{\tilde{P}}&=\frac{\mathrm{d}\log (\tilde{K}\Sigma^{\tilde{\gamma}})}{\mathrm{d} \log (R)} \\
\beta_{\tilde{S}}&=\frac{\mathrm{d}\log (\tilde{P}\Sigma^{-\tilde{\gamma}})}{\mathrm{d} \log (R)}=\beta_{\tilde{P}}-\tilde{\gamma}\beta_{\Sigma}.
\end{align}
Using this, we find a vertically integrated, radial Brunt V\"ais\"al\"a frequency
\begin{equation}
\tilde{N}_R^2=\frac{1}{\tilde{\gamma}}\chi^2 \beta_{\tilde{S}} \beta_{\tilde{P}} \Omega^2
\end{equation}
where $\chi=H/R$ represents the disk's aspect ratio with respect to the local pressure scale height $H$.

The growth rate for the SBI was derived by \cite{Lesur_Papaloizou_2010} and rewritten to a more easily applicable form by \cite{Beutel_2012} as  
\begin{equation}
\Gamma_{\mathrm{SBI}}\approx -\frac{4 \tilde{N}_R^2}{\omega (1+\chi^2)} \left(\frac{\tilde{\gamma} \omega \tau_{\mathrm{relax}}}{1+(\tilde{\gamma}\omega \tau_{\mathrm{relax}})^2}\right).
\label{SBI_growth}
\end{equation}
In order to get an estimate for a vortex' maximum growth rate, we set $\omega \tau_{\mathrm{relax}}=1$ and determined its angular velocity $\omega$ with the relation by \cite{Goodman_1987} 
\begin{equation}
\omega=\Omega \sqrt{\frac{3}{\chi^2 - 1}}.
\end{equation}
We choose the aspect ratio to be $\chi=4$, which is a reasonable value for large scale $R-\phi$ vortices \citep{Manger_2018}.
Plugging these assumptions into \autoref{SBI_growth}, leads to an approximate growth rate of
\begin{equation}
\Gamma_{\mathrm{SBI}}\approx -\frac{\tilde{N}_R^2}{4 \Omega}.
\end{equation}
The instability / amplification mechanism is the same as for the COS, see again Fig.~\ref{COSmech}.

\begin{figure}[ht] 
\centering
\includegraphics[width=1.0\textwidth]{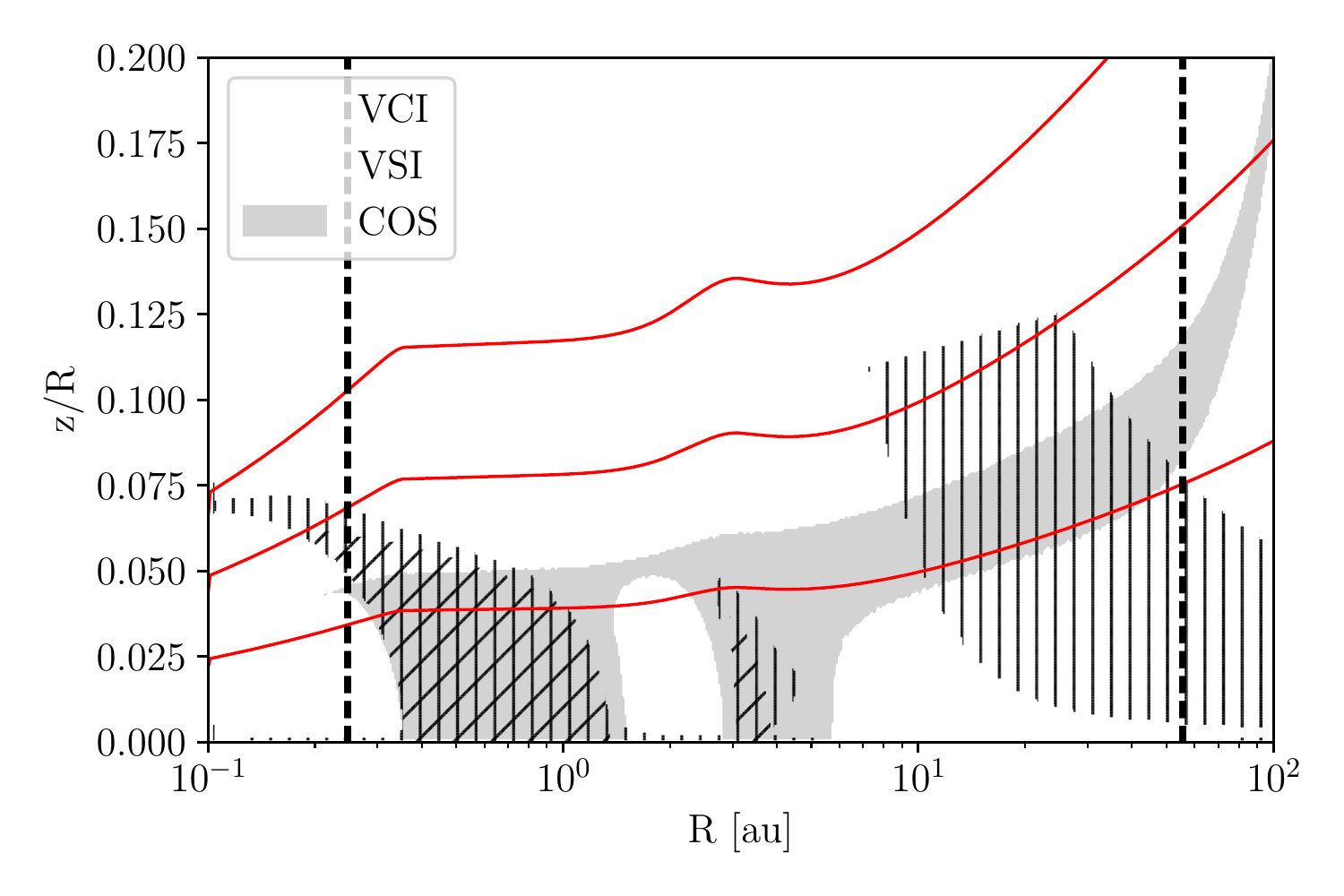}
\caption{Stability Map for a disk with $M_{\mathrm{disk}}=0.1\,M_{\mathrm{sun}}$ and $\alpha=10^{-3}$ around a solar mass star \citep{Pfeil2017}. Red lines indicate 1, 2 and 3 pressure scale heights. The baroclinic growth of vortices (SBI) can occur between the two thick dashed lines. VCI refers to Vertical Convective Instability (vertical buoyancy).}
\label{map}
\end{figure}

\subsection{Zombie Vortitces}
Another hydrodynamical effect creating vortices in disks would be the so called \textit{zombie vortex instability} (ZVI) \citep{2015ApJ...808...87M, 2016ApJ...833..148M}, which depends on the formation of critical layers in the disk, which communicate via waves in a vertically stably stratified disk. The precise conditions for the occurrence of this instability in realistic disks are still under discussion \citep{2016MNRAS.462.4549L}. This instability does not have a linear growth phase, in contrast to the other instabilities we discussed in the sections above, therefor studies of the ZVI mostly rely on numerical experiments. 
\section{Conclusion on Instabilities in Protoplanetary Disks}
We can conclude that several hydro-dynamical instabilities are potentially operating in protoplanetary disks. The discussed instabilities all depend on the thermodynamic stratification of the disk and on the local rate of thermal relaxation. 
Fig.~\ref{map} brings these findings together in the form of a map that constraints the spatial extent of hydro-dynamically unstable regions for: the Vertical Shear Instability (VSI), the Convective Overstability (COS), and the amplification of vortices via the Subcritical Baroclinic Instability (SBI).
For this map, \cite{Pfeil_2017} used an 1+1 dimensional, steady state accretion disk model, including the effects of stellar irradiation, viscous self-heating and flux-limited radiative transfer. This allows for a realistic determination of the local thermal relaxation rate, as well as for an investigation of the influence of the model its parameters, like stellar mass, disk mass and the viscosity parameter $\alpha$ on the conditions for instability.
The results imply that passively irradiated disks with $\alpha\lesssim 10^{-4}$ do only have VSI and COS susceptible zones at radial distances $\gtrsim \SI{10}{\bf AU}$ and about one pressure scale height above the midplane.
Vortex amplification via SBI nevertheless operates even for this very low viscous heating rates. 
As soon as $\alpha \gtrsim 10^{-4}$ is reached, VSI and COS become active and remain active  down to radial distances of even $\sim \SI{1}{\bf AU}$. Here, despite relatively long thermal relaxation times, VSI can actually operate even close to the midplane. This is because of the vertically adiabatic stratification of the viscously heated disk. 
The growth of all these considered instabilities (VSI, COS, and SBI) is favored for lower stellar masses and larger disk masses.

To conclude, these instabilities create moderate $\alpha$-stresses leading to viscous heating of the disk. They are found to operate right in the planet forming regions of protoplanetary disks and thus need to be considered when modeling disk dynamics.

\section{Structure Formation from Turbulence -- On Zonal Flows and Vortices in Conjunction with a Turbulent Cascade}
Turbulence is defined by velocity fluctuations that have a finite correlation length and correlation time. The first is a measure for the self-similarity over spatial distance, the latter for self-similarity over time. This means, turbulence is an unordered, time-varying velocity field, without any recognizable pattern. This comes from the fact that turbulent flows are strongly dissipative and thus quickly loose their 'memory'. Non-driven turbulence must therefor decay on a rate largely independent from the viscosity of the medium. At least as long as the dissipation scale is small enough with respect to the energy containing scales.

This is typically what is found for three-dimensional, isotropic turbulence within laboratory studies of driven or decaying turbulence. In these experiments, energy is typically injected at some length scale, that is called the \textit{integral length scale}, which can for instance be the size of an airplane wing. The induced eddies then trickle down to smaller and smaller eddy scales. This happens via the decaying of vortices under a constant energy transfer rate, until molecular viscosity is dissipating the turbulent energy, which naturally happens on the smallest possible scales, see Kolmogorov turbulence cascade in literature. 

If now the studied system is two-dimensional by construction, e.g. a thin water layer or earth its atmosphere, or the system is quickly spinning, e.g. planetary atmospheres, accretion disks, spinning laboratory flows, then turbulent energy cannot propagate to smaller scales. The reason lies in the energy transfer mechanism itself, as it involves the destruction of vortices in order to form smaller ones. This mechanism is called \textit{vortex stretching} and is inherently three-dimensional. Even if rotating fluids might be vertically thick with comparison to the considered scales, then global rotation still modifies the dynamical equations by introducing conservation of angular momentum. This means that the Coriolis forces start to dominate over the inertial forces. In other words, the small and fast rotating vortices do not notice the global rotation of a system, but large and slow vortices do. For example hurricanes are strongly affected by the rotation of the earth, but not so much are the small scale laboratory studies of turbulence affected, which naturally reside on the rotating earth, too.

For accretion disks this means that if any of the above mentioned instabilities inject kinetic energy with velocity $u$ at a length scale $l$, the Coriolis forces decide whether this energy can actually cascade down or not. The ratio of the perturbation frequency $\sim u/l$ to the rotation frequency of the system (disk) $\Omega$ defines a so called Rossby Number
\begin{equation}
\mathrm{Ro}=\frac{u}{l \Omega}\, ,
\end{equation}
which relates inertia against Coriolis forces. If the Rossby number is larger than unity, the eddy is in the 3-d Kolmogorov regime, but if the number is smaller than unity then one is in a 2-d regime. As explained, in the latter situation the energy cannot cascade down to smaller eddy sizes and thus the eddies can only merge into larger structures, i.e. (larger vortices), leading to an upward cascade, called 'inverse' cascade. 

\begin{figure}\centering
\includegraphics[scale=.65]{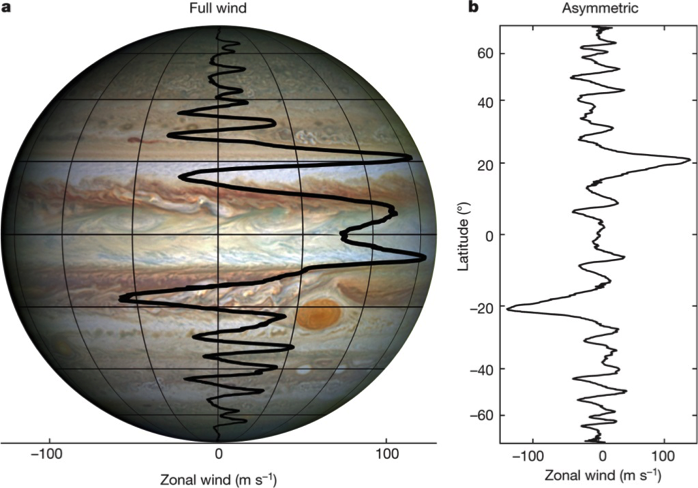}
\caption{Jupiter's asymmetric zonal velocity field. They are a result of perturbations that are driven on small scales and then upward cascade their energy. Note, that the big red spot is a vortex that grows by merging with smaller vortices and does not get destroyed over a very long period. Source: \cite{Kaspi2018}
}
\label{fig:Jupiter}
\end{figure}

This is the fundamental formation principle of large scale flow features in planetary atmospheres, starting from high and low pressure regions in Earth its atmosphere (cyclones and anti-cyclones), which find their counterpart in Jupiter and Saturn its atmospheres (see Fig.~\ref{fig:Jupiter}). As well as in Zonal-Flows, like the Jet-streams on Earth and likewise to the banded surface structure of gas giants, see Fig.~\ref{fig:Jupiter}. These zonal flows are axisymmetric winds, rotating faster or slower than the actual local rotation of the spinning body.

Accretion disks have much in common with planetary atmospheres. They both are spinning bodies with a typically low Rossby numbers, and if now turbulence is injected at a certain scale, independent on process and strength, then one will observe structure formation. The emergence of zonal flows and vortices is regularly reported for a wide range of magnetic and non-magnetic numerical simulations of accretion disks provided that the instabilities are weak. Weak means here that the created velocities are affected by rotation.
It is interesting to note, that a fully developed MRI is likely to disrupt the forming coherent structures,
making the formation and survival of vortices indeed less likely. Nevertheless, especially non-ideal MHD simulations find indeed the formation of vortices.

In past years of operation, ALMA and the VLA regularly observe indications for such vortices in protoplanetary disks \citep{2013Sci...340.1199V,2016ApJ...821L..16C}, yet this only proves that rotation is dominating their dynamics - it does not reveal the initial source of energy that led to the structure formation. The only instability that could have a direct influence on vortices, is the SBI, which could be able to amplify already existing vortices. Even though the instability cannot directly form them, due to the subcritical nature of it. This is similar to the process of energy being pumped into hurricanes from the release of latent heat in the heavy rain-falls in the hurricane. Percipitation is not creating the cyclone but it is stabilizing and even accelerating it, thus has a similar function as the SBI can have for vortices in accretion disks.

\section{Particle Trapping in Zonal Flows and Vortices}

In our introduction we already described that a particle disk, like Saturns rings, does behave differently from a gaseous disk.
A pure gas disk would orbit sub-Keplerian and be vertically thick, whereas a particle disk would orbit with Keplerian velocity and be vertically thin.
This arises because of the non-existing gas pressure support in a particle disk. The difference in both disk types can be explained by the fact that the rms-velocities of particles are much smaller than what the thermal velocity of the gas molecules would be.
If one is now combining both these components in a single coupled system, interesting behavior arise because of this different equilibrium states of the two components, as is the case in protoplanetary disks.

\subsection{Particles in a gas dominated disk}
Initially a protoplanetary disk can be seen as well mixed with small grains, leading to homogeneous dust-to-gas ratio of about $1:100$. 
At low dust-to-gas ratios the dynamics of the dust can be treated by studying individual particles. No collective effects will occur. The equation of individual dust grain is then given by the gravitational attraction from the star and by friction with the gas. Frictional forces are proportional to the relative velocity between gas and particle $\delta v = v_{\rm gas} - v_{\rm particle}$.
In the following we use the convention that $v = v_{\rm particle}$ and $u = v_{\rm gas}$ as introduced by \cite{1986Icar...67..375N}.

Friction is also a function of the cross-section and shape of the particle, e.g. its aerodynamic properties. Furthermore one has to distinguish between particles smaller than the mean-free-path of the gas (free molecular regime) and ones larger. These two regimes are called the Epstein regime and the Stokes regime, the latter when molecules act collectively as a fluid, because their mean-free-path is much smaller than the dimensions of the particles. We refer to \cite{1977MNRAS.180...57W} for a detailed description of the various regimes. For our purposes, we use the concept of a friction time $\tau_f$, which expresses the frictional acceleration of a dust grain to be linear proportional to the relative velocity, which is a good assumption for the expected small velocities in the solar nebula, both in the Epstein as well as in the Stokes regime. Then we can write
\begin{equation}
\partial_t \delta v = -\frac{\delta v}{\tau_f}.
\end{equation}
Thus, for an otherwise force free system, particles would be accelerated to the gas velocity on the e-folding time $\tau_f$. If on the other hand an additional force is acting on the particle, like gravitational acceleration $g$, then
\begin{equation}
\partial_t \delta v = -\frac{\delta v}{\tau_f} + g.
\end{equation}
For the equilibrium state ($\partial_t \delta v = 0$) the particle will move at the \textit{terminal velocity} given by
\begin{equation}
\delta v = \tau_f g,
\end{equation}
which is for instance a constant sedimentation velocity of a grain in a stratified atmosphere.
Additional forces, like self-gravity of the disk or coordinate effects, like Centrifugal and Coriolis accelerations also have to be included here, which can
make the determination of the dust behavior in a turbulent protoplanetary disk a complicated problem.

But, if we focus on smaller particles, with a friction time smaller than the dynamical time scales of the disk turbulence, then we can again use a terminal velocity ansatz. If we compare the dynamic equation describing the motion of a Lagrangian gas parcel under external and fictional forces, respectively accelerations $f$,
\begin{equation}
\partial_t u = -\frac{1}{\rho} \nabla p + f 
\label{gasdyn}
\end{equation}
with that of a particle
\begin{equation}
\partial_t v =  -\frac{\delta v}{\tau_f} + f,
\end{equation}
we can see that we can make the term $f$ disappear by subtracting
both equations and assume that $\partial_t (u - v) \ll \frac{u - v}{\tau_f}$, e.g. our argument that dynamical timescales are longer than the friction time.
Then we receive the handy expression:
\begin{equation}
v = u  + \frac{\tau_f}{\rho} \nabla p.
\label{eq:dustFollowsP}
\end{equation}
This means that small particles move on first order with the gas plus a small correction proportional to the local pressure gradient. This explains sedimentation to the midplane as much as radial drift towards the star. Also the trapping in local pressure maxima like in zonal flows and anti-cyclonic vortices can be understood in this fashion. All those long lived flow features must have an equilibrium between gravity, Coriolis and centrifugal acceleration together with the gas pressure gradient. Thus, their center must be a local pressure maximum. This is how particle traps work, and as one can see, the strength of the trapping depends on the strength of the pressure bump and on the friction timescale $\tau_f$. 

The necessity of these traps is discussed by Andrews and Birnstiel in their chapter 'Dust Evolution in Protoplanetary Disks'. In short, the problem in a disk without traps is that collisional bouncing and fragmentation, and the fast particle inward drift prevents the formation of any objects larger than a few centimeters, let alone planetesimals. As described in that chapter and Eq.~\ref{eq:dustFollowsP}, dust follows the pressure gradient in the disk, be it by sedimenting to the disk mid-plane, drifting radially inward or getting trapped in local pressure maxima of zonal flows \citep{Whipple1964,Dittrich2013} and vortices \citep{BargeSommeria1995, Raettig2015}.

One finds the radial drift rate for small objects (coupling time shorter than a local Keplerian orbit) is proportional to the size of the particles \citep{Weidenschilling1987}. Thus, we define \textit{pebbles} as the largest particles, that dominate the radial mass flux of solid material, whereas \textit{dust} is the smaller, non-drifting solid material. The radial flux in pebbles is then the key process in forming planetesimals, and ulimately regulating their formation rate \citep{Birnstiel2012}.

\section{The Dust Cloud Collapse Picture by Safronov and Goldreich \& Ward}
The gravitational collapse of local pebble concentrations in the solar nebula is meanwhile accepted to be an efficient route to form planetesimals, see the chapter by Armitage. The work by \cite{2006ApJ...636.1121J, 2007Natur.448.1022J, 2011A&A...529A..62J} demonstrated that pebble do concentrate in zonal flows, which were created in an MRI unstable accretion disk flow, were indeed concentrating beyond Hill density, and by that were gravitationally bound, instead of hydro-dynamically.

It should be stressed out that the gravitational collapse of a pebble layer in a protoplanetary disk is actually the original planetesimal formation idea, see the groundbreaking work by \cite{Safronov1972} and \cite{Goldreich1973}. But, \cite{Weidenschilling1987} found that the drag force feedback from pebbles residing in the mid-plane onto the gas actually leads to turbulence which is vertically diffusing the pebbles and by that preventing a final gravitational collapse to planetesimals. As we will see, the situation in particle traps is by todays understanding quiet similar but leads to a size threshold for collapsing particle clouds.

\subsection{Dust-to-gas ratios on the order of one: transition to a particle dominated disks}

Eq.~\ref{gasdyn} neglects the feedback from the particle friction on the gas and as soon as there is equal amount of dust $\rho_d$ and gas $\rho$ one has to write instead:
\begin{equation}
\partial_t u = -\frac{1}{\rho} \nabla p + \frac{\rho_d}{\rho}\frac{\delta v}{\tau_f}+ f .
\label{gasdyn2}
\end{equation}
This system now leads to a new equilibrium rotation profiles of the dust and gas components and also to an additional radial gas flux outward, which attributes for the angular momentum loss by the inward drifting particles.

\cite{1986Icar...67..375N} derived this equilibrium pattern for a given normalized radial pressure gradient $\eta = -\frac{1}{2\rho} \nabla p R \Omega^2$ and a dust-to-gas ratio of $\eps = \frac{\rho_d}{\rho}$:
\begin{equation}
\label{eq:NakagawaGas}
u_r = 2 \eps \stokes \lambda \eta\vk \qquad \text{and} \qquad u_\varphi = - \left(1 + \eps + \stokes^2 \right) \lambda \eta \vk\qquad \t{(Gas)}
\end{equation}
\begin{equation}
\label{eq:NakagawaDust}
v_r =   - 2\stokes\lambda \eta\vk  \qquad \text{and} \qquad v_\varphi = - \left(1 + \eps \right) \lambda \eta\vk \qquad \t{(Dust)}
\end{equation}
$\stokes = \Omega \tau_f$ is here the normalized friction time and we use the notation:
\begin{equation}
\lambda \defEq \frac{1}{\left(1+\eps \right)^2+ \stokes^2}\,.
\end{equation}
Now that we have an interesting equilibrium state for a given dust-to-gas ratio $\eps$, one can investigate the stability of the system for perturbations in $\eps$. If, for instance, $\eps \gg 1$ then in the $\phi$-component of the gas velocity, see Eq.~\ref{eq:NakagawaGas}, the $\lambda$ approaches zero and the gas disk in no longer sub-Keplerian.
This lead to the discovery of the Kelvin-Helmholtz Instability (KHI) due to the vertically sedimenting dust, which is forming a dens dust sub-disk in the mid-plane and then to a strong vertical shear between this Keplerian dust disk and the sub-Keplerian gas disk \cite{Weidenschilling1987}. The KHI then stops further sedimentation via turbulent diffusion and no gravitational collapse should happen.

Another conclusion from Eq.~\ref{eq:NakagawaDust} is that for dust-to-gas ratios above one, the radial drift velocity of the dust decreases proportional to $\eps^{-2}$. Thus, the denser a particle swarm becomes the slower it drifts radially.
This is the basic effect of the \textit{Streaming Instability} for a radially inward drifting particle swarm. 
It should be noted here that very compacted particle heaps actually have a cumulative Stokes number that is different from the individual particle Stokes number and dynamics might become different again.

The \textit{streaming instability} (SI), as described in \citep{2005ApJ...620..459Y,Johansen2007,Youdin2007}, starts around dust-to-gas ratio of unity. A simplified version of their derivation can be found in \cite{Jacquet2011} and lately \cite{Squire2018SI,Squire2018} showed that the SI is actually part of a larger family of instabilities, which is called the resonant drag instability.

Sometimes the SI is compared to a traffic jam. Because pebble clumps with a higher dust-to-gas ratios drift slower, it is argued that the following clumps with a lower dust-to-gas ratio will collide with the slower moving clump leading to a further increase in density. But this not true, because the pebble drift is not linear unstable if one considers only the radial modes for the perturbation. In fact one needs two dimensions, typically radial and vertical direction to set up complicated modes that actually show the further increase of the dust-to-gas ratio as result of the SI \citep{Johansen2007,Youdin2007}.

\subsection{Pure Streaming Instability Case}
In the absence of zonal flows and vortices, pebbles sediment to the mid-plane, then KHI kicks in and further dynamics are generated by the SI, see \cite{Bai_2013}. But, for the expected global dust-to-gas ratios in the disk around a young star, self-gravity will not be triggered and no planetesimals can form as was explored by \cite{2009ApJ...704L..75J}. Only if the global dust-to-gas ratio is increased, for instance at the late stages of the disk evolution when photoevaporation removes gas from the disk \citep{2017ApJ...839...16C}, then self-gravity is strong enough to overcome the SI and planetesimals can form. But, this is clearly too late to form gas giants, but maybe comets. The conclusion here is that without particle traps no planetesimals will form early enough for planet formation. 

In the past, the capability of the SI to locally produce strong density fluctuations in the dust density has been seen as a beneficial process for planetesimal formation and has been studied in stratified simulations, e.g. by \cite{Yang2017} and \cite{Carrera2015}. Works by \cite{Johansen2015}, \cite{Schaefer2017} and \cite{Simon2016} tried to estimate an initial mass function for planetesimals in disks with an (unrealistic) high amount of pebbles and consequently produced to many large planetesimals. As noted in \cite{Klahr2015}, the streaming instability should actually be seen to hinder the dust mid-plane layer from collapsing \citep{Bai_2013} but also limits particle cloud collapse on small scales.
One has to conclude that there is no planetesimal formation 'by' the streaming instability, but instead it is rather the SI which regulates the onset of planetesimal formation. The initial size distribution then dependents on how much material gets concentrated locally in a trap and how strong the SI is, the latter is based on the particle size, the gas pressure gradient and dust-to-gas ratio.

Only if the dust-to-gas ratio is high enough, turbulent stirring from the particle feedback (aka streaming instability) can be overcome triggering gravitational collapse \cite{Johansen2010}.
The sufficient local dust accumulations needed for gravitational collapse to overcome the SI can be argued to be the result of dust trapping in vortices and zonal flows \cite{Klahr2015}. 

\section{And then: Planetesimal Collisions and Pebble accretion.}
As planetesimals start to form in the circumstellar disk they also start to collide among each other \cite{Safronov1972}, eventually growing to larger and larger objects.
This process is not directly controlled by the gas of the nebula, but turbulence with its potential density fluctuations on large scales 
can pump additional kinetic energy into the planetesimal population \cite{2007Icar..188..522O}. These collisions in conjunction with the accretion of additional pebbles from the disk \cite{Ormel2010}, is then the route to form  planetary embryos, but this is a different story to be told elsewhere in this section.

\section{Conclusion}
Turbulence was initially introduced to explain angular momentum transport in the circumstellar disk to allow for gas accretion through and from the disk onto the star, an observable effect that needs explanation. In the current state of research it is not clear how much angular momentum is in fact transported by turbulence and how much by winds form the disk \cite{1992ApJ...394..117P, 2013ApJ...769...76B,2015ApJ...801...84G}. Strong turbulence is also neither supported by observations \cite{2018ApJ...856..117F} nor could it be explained with our knowledge of magnetic and hydro-dynamic effects in disks. Yet, weak turbulence consistent with the so far studied instability mechanisms (both of magnetic and pure hydro nature) is consistent with the formation of large scale structures in disks around young stars, which are vital to trigger the formation of planetesimals via the trapping of pebbles and the gravitational collapse of pebble clouds, a fundamental step to form planetary systems.

\bibliographystyle{spbasicHBexo}  
\bibliography{HBexoTemplateBib} 

\end{document}